\documentclass[10pt]{article}
\usepackage{latexsym,graphicx}
\usepackage{subfigure}
\usepackage{amssymb}
\usepackage{amsmath}
\usepackage{amscd}
\usepackage{amsthm}
\usepackage[left=2cm,top=2.5cm,right=2.5cm,bottom=1.5cm]{geometry}
\begin{document}
\begin{center}
\large{\bf{Bianchi Type-$V$ cosmology in $f(R,T)$ gravity with $\Lambda(T)$}} \\
\vspace{10mm}
\normalsize{Nasr Ahmed$^{1,2}$ and Anirudh Pradhan${^3}$}\\
\vspace{5mm}
\normalsize{$^{1}$ Mathematics Department, Deanery of Academic Services, Taibah University, Ai-Madinah Al-Munawwarah, 
Saudi Arabia} \\
\vspace{5mm}
\normalsize{$^{2}$ Astronomy Department, National Research Institute of Astronomy and Geophysics, Helwan, Cairo, Egypt \\
\vspace{2mm}
E-mail: nasr.ahmed@nriag.sci.eg} \\
\vspace{5mm}
\normalsize{$^{3}$Department of Mathematics, Hindu Post-graduate College, 
Zamania-232 331, Ghazipur, India \\
\vspace{2mm}
E-mail: pradhan@iucaa.ernet.in} \\
\end{center}
\vspace{10mm}
\begin{abstract}
A new class of cosmological models in $f(R, T)$ modified theories of gravity proposed by Harko et al. (2011), has been investigated for a specific choice of $f(R, T) = f_{1}(R) + f_{2}(T)$ by considering time dependent deceleration 
parameter. The concept of time dependent deceleration parameter (DP) with some proper assumptions yield the average scale factor 
$a(t) = \sinh^{\frac{1}{n}}(\alpha t)$, where $n$ and $\alpha$ are positive constants. For $0 < n \leq 1$, this generates 
a class of accelerating models while for $n > 1$, the models of universe exhibit phase transition from early decelerating phase 
to present accelerating phase which is in good agreement with the results from recent astrophysical observations. Our intention 
is to reconstruct $f(R,T)$ models inspired by this special law for the deceleration parameter in connection with the theories of 
modified gravity. In the present study we consider the cosmological constant $\Lambda$ as a function of the trace of the stress 
energy-momentum-tensor, and dub such a model  ``$\Lambda(T)$ gravity'' where we have specified a certain form of $\Lambda(T)$. Such 
models may display better uniformity with the cosmological observations. The statefinder diagnostic pair $\{r,s\}$ parameter 
has been embraced to characterize different phases of the universe. We also discuss the physical consequences of the derived 
models. 
\end{abstract}
\smallskip
{\it Keywords}: Bianchi type-$V$ universe, Modified gravity, Variable deceleration parameter, Time dependent $\Lambda$-term  \\
{\it PACS number}: 04.50.Kd, 98.80.Es
\section{Introduction}
Despite the fact that Einstein's general relativity (EGR) is a triumphal success theory and the basis for the 
description of most of gravitational phenomena known to date, it fails to explain the recent discovery of the 
accelerating expansion of the universe. The recent observational data [1$-$13] strongly indicate that our universe 
is dominated by a component with negative pressure, dubbed as dark energy (DE), which constitutes with $\simeq 3/4$ 
of the critical density. In order to explain the nature of dark energy and the accelerated expansion, a variety of the 
theoretical models have been proposed in the literature such as quintessence, phantom energy, k-essence, tachyon, 
f-essence, chaplygin gas, etc. Among these different models of DE, the modified gravity theories are f(R) gravity 
[14$-$18] and Gauss-Bonnet gravity or f(G) gravity [19$-$23]. Another approach to modified gravity theory is so called 
f(T) gravity [24$-$26], where $T$ is scalar torsion. These modified gravities have recently been 
verified to explain the late-time accelerated expansion of the Universe. \\

One of the interesting and prospective versions of modified gravity theories is the $f(R, T)$ gravity proposed by 
Harko et al. \cite{ref27}. The $f(R, T)$ gravity models can explain the late time cosmic accelerated expansion of the 
Universe. Recently, Chaubey and Sukla \cite{ref28}, Adhav \cite{ref29}, Samanta \cite{ref30}, Reddy et al. [31$-$33]
have studied cosmological models in $f(R, T)$ gravity in different Bianchi type space-times. \\

In recent years Bianchi universes are playing important role in observational cosmology, since the WMAP data [34$-$36] 
seem to require an addition to the standard cosmological model with positive cosmological constant that bears a likeness 
to the Bianchi morphology [37$-$42]. According to this, the universe should reach to a slightly anisotropic special 
geometry in spite of the inflation, contrary to generic inflationary models [43$-$49] suggesting a non-trivial isotropization 
history of universe due to the presence of an anisotropic energy source. In order to explain the homogeneity and flatness of 
the presently observed Universe, it is usually assumed that this has undergone a period of exponential expansion [43, 45$-$47]. 
Mostly the expansion of the universe is described within the framework of the homogeneous and isotropic Friedman-Robertson-Walker 
(FRW) cosmology. The reasons for this are purely technical. The simplicity of the field equations and the existence of analytical 
solutions in most of the cases have justified this over simplification for the geometry of space-time. However, there are no 
compelling physical reasons to assume the former before the inflationary period. To drop the assumption of homogeneity would 
make the problem interactable, while the isotropy of the space is something that can be relaxed and leads to anisotropy. Several 
authors [50$-$54] have studied particular cases of anisotropic models and found that the scenario predicted by the FRW model stand 
essentially unchanged even when large anisotropies were present before the inflationary period. Thus, the anisotropic Bianchi models 
become of academical interest. Out of this, in the present study Bianchi type-V space-time is taken into consideration. \\

Motivated by the above discussions, in this paper, we propose to study cosmological model in $F(R, T)$ gravity in Bianchi type-$V$ 
space-time by considering $f(R, T) = f_{1}(R) + f_{2}(T)$. The outline of this paper is as follows: In Sect. $2$, gravitational field 
equation in the $f(R,T)$ modified gravity theory is established. Section $3$ deals with the basic equations and quadrature solutions. 
In Sect. $4$, the physical and geometric aspects of the model have been discussed. Section $5$ deals with the physical acceptability 
of the solutions. Finally, conclusions are summarized in the last Sect. $6$.  
\section{$f(R,T)$ modified gravity theory}
The theory suggests a modified gravity action given by
\begin{equation}
\label{eq1} S = \frac{1}{16\pi}\int{f(R,T)\sqrt{-g}d^{4}x} + \int{L_{m} \sqrt{-g}d^{4}x} ,
\end{equation}
where $f(R,T)$ is an arbitrary function of the Ricci scalar, $R$, and the trace $T$ of the stress-energy tensor 
of the matter, $T_{\mu \nu}$. $L_{m}$ is the matter Lagrangian density. The stress-energy tensor of matter is defined as
\begin{equation}
\label{eq2} T_{\mu \nu} = -\frac{2}{\sqrt{-g}}\frac{\delta{\sqrt{-g}L_{m}}}{\delta g^{\mu \nu}} ,
\end{equation}
and its trace by $T = g^{\mu \nu}T_{\mu \nu}$. Assuming that the Lagrangian density $L_{m}$ of matter depends only on the components of the 
metric tensor $g_{\mu \nu}$ and not on its derivatives leads to
\begin{equation}
\label{eq3} T_{\mu \nu} = g_{\mu \nu} L_{m} - 2\frac{\partial L_{m}}{\partial g^{\mu \nu}}.
\end{equation}
Here and in this subsection, we use the unit of $G = c = 1$. As past related studies, a theory whose Lagrangian density is
described by an arbitrary function of R and the Lagrangian density of matter as $F(R,L_{M})$ has been explored in Harko
and Lobo \cite{ref55}. Moreover, in Poplawski \cite{ref56} a theory in which the cosmological constant is written by a function
of the trace of the stress-energy tensor as $\Lambda(T)$ has been investigated. \\

Varying the action $S$ with respect to the metric tensor components $g^{\mu \nu}$, the gravitational field equations of 
$f(R,T)$ gravity is obtained as 
\[
 f_{R}(R,T)R_{\mu \nu}-\frac{1}{2} f(R,T)g_{\mu \nu}+(g_{\mu \nu} \Box  -\nabla_{\mu} \nabla_{\nu})f_{R}(R,T) 
\]
\begin{equation}
\label{eq4}
= 8\pi T_{\mu \nu} - f_{T}(R,T)T_{\mu \nu} - f_{T}(R,T)\Theta_{\mu\nu}. 
\end{equation}
with $\Theta_{\mu \nu} \equiv g^{ij}\left(\frac{\delta T_{ij}}{\delta g^{\mu \nu}}\right)$, which follows from the relation $\delta \left(\frac{g^{ij}T_{ij}}
{\delta g^{\mu \nu}}\right) = T_{\mu \nu} + \Theta_{\mu \nu}$ and $\Box = \nabla^{i}\nabla_{i}$, $f_{R}(R,T) \equiv \frac{\partial f(R,T)}{\partial R}$, 
$f_{T}(R,T) \equiv \frac{\partial f(R,T)}{\partial T}$ and $\nabla_i$ denotes the covariant derivative. The contraction of Eq. (\ref{eq4}) yields 
$ f_{R}(R,T)R  + 3 \Box f_{R}(R,T) - 2f(R,T) = (8\pi - f_{T}(R,T))T - f_{T}(R,T)\Theta$ with $\Theta \equiv g^{\mu \nu}\Theta_{\mu \nu}$. Combining Eq. 
(\ref{eq4}) and the contracted equation and eliminating the $\Box f_{R}(R,T)$ term from these equations, we obtain
\[
 f_{R}(R,T)\left(R_{\mu \nu} - \frac{1}{3}R g_{\mu \nu}\right) + \frac{1}{6}f(R,T)g_{\mu \nu}
\]
\[
 = (8\pi - f_{T}(R,T))\left(T_{\mu \nu} - \frac{1}{3}T g_{\mu \nu}\right)
\]
\begin{equation}
\label{eq5}
- f_{T}(R,T)\left(\Theta_{\mu \nu} - \frac{1}{3}\Theta g_{\mu \nu}\right) + \nabla_{\mu} \nabla_{\nu}f_{R}(R,T) .
\end{equation}
On the other hand, the covariant divergence of Eq. (\ref{eq1}) as well as the energy-momentum conservation law 
$\nabla^{\mu}\left[f_{R}(R,T) - \frac{1}{2}f(R,T)g_{\mu \nu} + (g_{\mu \nu} \Box - \nabla_{\mu}\nabla_{\mu})f_{R}(R,T)\right] 
\equiv 0$, which corresponds to the divergence of the left-hand side of Eq. (\ref{eq1}), we acquire the divergence of $T_{\mu \nu}$ 
as 
\begin{equation}
\label{eq6}
\nabla^{\mu}T_{\mu \nu} = \frac{f_{T}(R,T)}{8\pi - \frac{1}{2}f_{T}(R,T)}\biggl[(T_{\mu \nu} + \Theta_{\mu \nu})\nabla^{\mu}\ln f_{T}(R,T) 
+ \nabla^{\mu}\Theta_{\mu \nu}\Biggr].
\end{equation}
In addition, from $T_{\mu \nu} = g_{\mu \nu}L_{M} - 2\left(\frac{\partial{L_M}}{\partial{g^{\mu \nu}}}\right)$ we have
\begin{equation}
\label{eq7}
\frac{\delta T_{ij}}{\delta g^{\mu \nu}} = \left(\frac{\delta g_{ij}}{\delta g^{\mu \nu}} + \frac{1}{2}g_{ij}g_{\mu \nu}\right)L_{M} 
- \frac{1}{2}g_{ij} T_{\mu \nu} - 2\frac{\partial^{2}L_{M}}{\partial g^{\mu \nu}\partial g^{ij}}.
\end{equation}
Using the relation $\frac{\delta g_{ij}}{\delta g^{\mu \nu}} = - g_{i\gamma}g_{j \sigma}\delta^{\gamma \sigma}_{\mu \nu}$ with 
$\delta^{\gamma \sigma}_{\mu \nu} = \frac{\delta g^{\gamma \sigma}}{\delta g^{\mu \nu}}$, which follows from $g_{i\gamma}g^{\gamma j} = \delta^{j}_{i}$, 
we obtain $\Theta_{\mu \nu}$ as given by
 \begin{equation}
\label{eq8} \Theta_{\mu \nu} = -2T_{\mu \nu} + g_{\mu \nu}L_{m} - 2g^{ij}\frac{\partial^{2}L_{m}}{\partial g^{\mu\nu}
\partial g^{ij}}.         
\end{equation}
Provided that matter is regarded as a perfect fluid, the stress-energy tensor of the matter Lagrangian is given by
\begin{equation}
\label{eq9} T_{\mu\nu} = (\rho + p)u_{\mu}u_{\nu} - pg_{\mu \nu}, 
\end{equation}
where $u^{\mu} = (0,0,0,1)$ is the four velocity in the moving coordinates which satisfies the conditions $u^{\mu}u_{\nu} = 1$ 
and $u^{\mu}\nabla_{\nu}u_{\mu} = 0$. $\rho$ and $p$ are the energy density and pressure of the fluid respectively. With the 
use of equation (\ref{eq8}), we obtain
\begin{equation}
\label{eq10} \Theta_{\mu \nu} = -2T_{\mu\nu}  - p g_{\mu\nu}.
\end{equation}
Since the field equations in $f(R,T)$ gravity also depend on the physical nature of the matter field (through the 
tensor $\Theta_{\mu\nu}$), several theoretical models can be obtained for each choice of $f$. Three explicit specification 
of the functional form of $f$ has been considered in Harko et al. \cite{ref27}
\begin{displaymath}
   f(R,T) = \left\{
  \begin{array}{lr}
       R + 2f(T)\\
       f_{1}(R) + f_{2}(T)\\
       f_{1}(R) + f_{2}(R)f_{3}(T)
     \end{array}
   \right.
\end{displaymath}
The cosmological consequences for the class  $f(R, T) = R + 2f(T)$ have been recently discussed in details by many authors [28$-$33, 57]. 
Recently, Shamir et al. \cite{ref58} and Chaubey \& Sukla \cite{ref28} have discussed Bianchi type-I \& V and a general class of Bianchi 
models respectively in $F(R,T)$ gravity by considering $f(R,T) = R + 2f(T)$. In this paper we are considering the cosmological consequences 
of the class for which $f(R,T) = f_{1}(R) + f_{2}(T)$. Our derived cosmological model is totally different and new from that of other authors 
mentioned here. So far the physically important cosmological term $\Lambda$ which is a candidate for dark energy remains less attended. So, 
our derived model may lead to better understanding of the characteristic of Bianchi type-V models. \\

The gravitational field equation (\ref{eq4}) becomes
\[
 f^{'}_{1}(R)R_{\mu \nu} -\frac{1}{2} f_{1}(R)g_{\mu \nu} + (g_{\mu \nu} \Box  -\nabla_{\mu} \nabla_{\nu})f^{'}_{1}(R) =
\]
\begin{equation}
\label{eq11}
 8\pi T_{\mu \nu} + f^{'}_{2}(T)T_{\mu \nu} + \left(f^{'}_{2}(T)p + \frac{1}{2}f_{2}(T)\right) g_{\mu\nu}, 
\end{equation}
where the prime denotes differentiation with respect to the argument. the field equations of the standard $f(R)$ gravity can 
be recovered for $p = 0$ (the dust case) and $f_{2}(T) = 0$. We consider a particular form of the functions $f_{1}(R) = \lambda_{1}R$ 
and $f_{2}(T) = \lambda_{2}T$ where $\lambda_{1}$ and $\lambda_{2}$ are arbitrary parameters. In this article we take 
$\lambda_{1} = \lambda_{2} = \lambda$ so that $f(R,T) = \lambda (R + T)$. \\
 
Equation (\ref{eq11}) can now be rewritten as
\[
 \lambda R_{\mu \nu}-\frac{1}{2} \lambda (R + T)g_{\mu \nu} + (g_{\mu \nu} \Box - \nabla_{\mu} \nabla_{\nu})\lambda
\]
\begin{equation}
\label{eq12}
= 8\pi T_{\mu \nu} -\lambda T_{\mu \nu} + \lambda (2T_{\mu \nu} + pg_{\mu \nu}).
\end{equation}
Setting $(g_{\mu \nu} \Box  -\nabla_{\mu} \nabla_{\nu})\lambda = 0$, we get
\begin{equation}
\label{eq13} \lambda G_{\mu\nu} = 8\pi T_{\mu \nu} + \lambda T_{\mu \nu} + \left(\lambda p + \frac{1}{2}\lambda T\right) g_{\mu\nu},
\end{equation}
where $G_{\mu\nu} = R_{\mu\nu} - \frac{1}{2}g_{\mu\nu}R$ is the Einstein tensor. This could be rearranged as
\begin{equation}
\label{eq14} G_{\mu\nu} - \left(p + \frac{1}{2} T\right) g_{\mu\nu} = \frac{8\pi + \lambda}{\lambda} T_{\mu \nu}.      
\end{equation}
Recalling Einstein equations with cosmological constant
\begin{equation}
\label{eq15} G_{\mu\nu}-\Lambda g_{\mu\nu}=-8\pi T_{\mu \nu}. 
\end{equation}
We choose a negative small value for the arbitrary $\lambda$ so that we have the same sign of the RHS of (\ref{eq13}) 
and (\ref{eq14}), we keep this choice of $\lambda$ throughout. The term $\left(p + \frac{1}{2} T\right)$  
can now be regarded as a cosmological constant. Hence we write
\begin{equation}
\label{eq16} \Lambda \equiv \Lambda(T) = p + \frac{1}{2}T.
\end{equation}
The dependence of the cosmological constant $\Lambda$ on the trace of the energy momentum tensor $T$ has been proposed 
before by Poplawski \cite{ref56} where the cosmological constant in the gravitational Lagrangian is a function of the trace of the 
energy-momentum tensor, and consequently the model was denoted ``$\Lambda(T)$ gravity''. It was argued that recent cosmological 
data favour a variable cosmological constant which is consistent with $\Lambda(T)$ gravity without the need to specify an exact 
form of the function $\Lambda(T)$. $\Lambda(T)$ gravity is more general than the Palatini $f(R)$ and could be reduced to it if 
the pressure of matter is neglected (Magnano\cite{ref59}, Poplawski \cite{ref60,ref61}). Considering the perfect fluid case, the trace 
$T = -3p + \rho$, for our model. Hence Eq. (\ref{eq16}) reduces to 
\begin{equation}
\label{eq17} \Lambda = \frac{1}{2}(\rho - p). 
\end{equation}

\section{The basic equations and quadrature solutions}
Modern observations (WMAP data for example) indicate that the universe is not completely symmetric [34$-$36]. From that 
point of view Bianchi models (which represents spatially homogeneous and anisotropic spaces) are more appropriate in describing 
the universe as it has less symmetry than the standard FRW models. We use the following metric of general class of Bianchi 
cosmological model
\begin{equation}
\label{eq18} ds^{2} = dt^{2} - A^{2}dx^{2} - e^{-2 \beta x}[B^{2}dy^{2} - C^{2}dz^{2}],
\end{equation} 
where $\beta$ is a constant and the functions $A(t)$, $B(t)$ and $C(t)$ are the three anisotropic directions of expansion in normal three 
dimensional space. Those three functions are equal in FRW models due to the radial symmetry and so we have only one function 
$a(t)$ there. The average scale factor $a$, the spatial volume $V$ and the average Hubble's parameter 
$H$ are defined as
\begin{equation}
\label{eq19} a = (ABC)^{\frac{1}{3}}, 
\end{equation} 
\begin{equation}
\label{eq20} V = a^{3} = ABC,
\end{equation} 
and 
\begin{equation}
\label{eq21} H = \frac{1}{3}(H_{1} + H_{2} + H_{3}),  
\end{equation} 
respectively with $H_{1} = \frac{{\dot{A}}}{A}$, $H_{2} = \frac{{\dot{B}}}{B}$ and $H_{3} = \frac{{\dot{C}}}{C}$. 
Here and elsewhere the dot denotes differentiation with respect to cosmic time $t$. From  Eqs. (\ref{eq1}), (\ref{eq2}) and (\ref{eq3})
we get
\begin{equation}
\label{eq22} H = \frac{1}{3}\frac{\dot{V}}{V} = \frac{1}{3}\left(\frac{\dot{A}}{A} + \frac{\dot{B}}{B} + 
\frac{\dot{C}}{C}\right).
\end{equation}

Now the cosmological equations for the energy momentum tensor (\ref{eq9}) and the metric (\ref{eq18}) are
\begin{equation}
\label{eq23}
 \frac{\dot{B}\dot{C}}{AC} + \frac{\ddot{B}}{B} + \frac{\ddot{C}}{C} - \frac{\beta^{2}}{A^{2}} 
=  \left(\frac{8\pi + \lambda}{\lambda}\right)p - \Lambda,
\end{equation}
 \begin{equation}
\label{eq24}
 \frac{\dot{A}\dot{C}}{AC} + \frac{\ddot{A}}{A} + \frac{\ddot{C}}{C} - \frac{\beta^{2}}{A^{2}} 
=  \left(\frac{8\pi + \lambda}{\lambda}\right)p - \Lambda,
\end{equation}
\begin{equation}
\label{eq25}
 \frac{\dot{A}\dot{B}}{AB} + \frac{\ddot{A}}{A} + \frac{\ddot{B}}{B} -\frac{\beta^{2}}{A^{2}} 
=  \left(\frac{8\pi + \lambda}{\lambda}\right)p - \Lambda.
\end{equation}
\begin{equation}
\label{eq26}
\frac{\dot{A}\dot{B}}{AB} + \frac{\dot{A}\dot{C}}{AC} + \frac{\dot{B}\dot{C}}{BC} - 
\frac{3\beta^{2}}{A^{2}} = - \left(\frac{8\pi +\lambda}{\lambda}\right)\rho - \Lambda,
\end{equation}
\begin{equation}
\label{eq27}
2\frac{\dot{A}}{A} - \frac{\dot{B}}{B} - \frac{\dot{C}}{C} = 0. 
\end{equation}
In the next proceeding of the paper we take constant $(\beta)$ as unity without any loss of generality. \\

Integrating Eq. (\ref{eq27}) and absorbing the integration constant into $B$ or $C$, we obtain
\begin{equation}
\label{eq28}
A^{2} = BC,
\end{equation}
without any loss of generality. Subtracting (\ref{eq23}) from (\ref{eq24}), (\ref{eq23}) from (\ref{eq25}), and (\ref{eq24}) from 
(\ref{eq25}) and taking second integral of each, we obtain the following three relations respectively:
\begin{equation}
\label{eq29} \frac{A}{B} = d_{1}\exp{\left(k_{1}\int{\frac{dt}{a^{3}}}\right)},
\end{equation}
\begin{equation}
\label{eq30} \frac{A}{C} = d_{2}\exp{\left(k_{2}\int{\frac{dt}{a^{3}}}\right)},
\end{equation}
and
\begin{equation}
\label{eq31} \frac{B}{C} = d_{3}\exp{\left(k_{3}\int{\frac{dt}{a^{3}}}\right)},
\end{equation}
where $d_{1}$, $d_{2}$, $d_{3}$, $k_{1}$, $k_{2}$ and $k_{3}$ are constants of integration. 
Finally, using $a = (ABC)^{\frac{1}{3}}$, we write the metric functions from  (\ref{eq29})$-$(\ref{eq31}) in explicit form as
\begin{equation}
\label{eq32}
A(t) = l_{1} a \exp \left(m_{1} \int {a^{-3}dt} \right),
\end{equation}
\begin{equation}
\label{eq33}
B(t) = l_{2} a \exp \left(m_{2} \int {a^{-3}dt} \right),
\end{equation}
\begin{equation}
\label{eq34}
C(t) = l_{3} a \exp \left(m_{3} \int {a^{-3}dt} \right),
\end{equation}
where 
\begin{equation}
\label{eq35}
l_{1} = \sqrt[3]{d_{1}d_{2}}, \indent l_{2} = \sqrt[3]{d_{1}^{-1}d_{3}}, \indent l_{3} = 
\sqrt[3]{(d_{2}d_{3})^{-1}},
\end{equation}
and
\begin{equation}
\label{eq36}
m_{1} = \frac{k_{1} + k_{2}}{3}, \indent  m_{2} = \frac{k_{3} - k_{1}}{3}, \indent  m_{3} = 
\frac{-(k_{2} + k_{3})}{3},
\end{equation}
The constants $m_{1}, m_{2}, m_{3}$ and $l_{1}, l_{2}, l_{3}$ satisfy the fallowing two relations:
\begin{equation}
\label{eq37}
m_{1} + m_{2} + m_{3} = 0, \; \; \; l_{1}l_{2}l_{3} = 1.
\end{equation}
Substituting Eq. (\ref{eq28}) in Eqs. (\ref{eq32})$-$ (\ref{eq34}), we obtain
\begin{equation}
\label{eq38}
 l_{1} = 1, \; l_{2} = l_{3}^{-1} = c_{1}, \; \; m_{1} = 0, \; m_{2} = -m_{3} = c_{2}, 
\end{equation}
where $c_{1}$ and $ c_{2} $ are constants. Again, substituting Eq. (\ref{eq38}) in Eqs. (\ref{eq32})$-$(\ref{eq34}), the 
quadrature form of the metric functions in terms of average scale factor a can be written as
\begin{equation}
\label{eq39} A(t) = a,
\end{equation}
\begin{equation}
\label{eq40} B(t) = c_{1} a\exp{\left(c_{2} \int{\frac{dt}{a^{3}}}\right)},
\end{equation}
\begin{equation}
\label{eq41} C(t) = \frac{a}{c_{1}}\exp{\left(-c_{2} \int{\frac{dt}{a^{3}}}\right)},
\end{equation}

Following Pradhan \cite{ref62}, Pradhan et al. \cite{ref63} and Chawla et al. \cite{ref64}, the modified gravity field equations 
can be solved by considering the time-dependent deceleration parameter which yield the average scale factor as
\begin{equation}
\label{eq42} a(t) = (\sinh(\alpha t))^{\frac{1}{n}}.
\end{equation}
The derivation and the motivation to choose such scale factor has already been described in details by Pradhan \cite{ref62}. 
This relation (\ref{eq42}) generalizes the value of scale factor obtained by Pradhan et al. \cite{ref65,ref66} in connection 
with the study of dark energy models in Bianchi type-$VI_{0}$ space-time and cosmological models with variable $q$ and 
$\Lambda$-term respectively. In literature it is common to use a constant deceleration parameter [28$-$33, 67$-$69], as it 
duly gives a power law for metric function or corresponding quantity. For the observational motivation of such time dependent 
DP see (Riess et al. \cite{ref4,ref5}; Perlmutter et al. \cite{ref1}; Tonry et al. \cite{ref9}; Clocchiatti et al. \cite{ref10}). \\

Eqs. (\ref{eq39})$-$(\ref{eq41}) with the help of (\ref{eq42}), we obtain the metric functions as
\begin{equation}
\label{eq43} A(t) = \sinh^{\frac{1}{n}}(\alpha t)),
\end{equation}
\begin{equation}
\label{eq44} B(t) = c_{1} \sinh^{\frac{1}{n}}(\alpha t)\exp{\left[\frac{c_{2}(-1)^{\frac{n+3}{2n}}}{2\alpha}
\cosh(\alpha t) F(t)\right]},
\end{equation}
\begin{equation}
\label{eq45} C(t) = \frac{1}{c_{1}}\sinh^{\frac{1}{n}}(\alpha t)\exp{\left[-\frac{c_{2}(-1)^{\frac{(n+3)}{2n}}}{2\alpha}
\cosh(\alpha t) F(t)\right]},
\end{equation}
where
\begin{equation}
\label{eq46} 
F(t) = 1 + \frac{1}{6}\left(1+\frac{3}{n}\right)\cosh^{2}(\alpha t) + \frac{3}{40}\left(1+\frac{3}{n}\right)
\left(1+\frac{1}{n}\right)\cosh^{4}(\alpha t) + o[\cosh (\alpha t)]^{6}.
\end{equation}
\begin{figure}
 \centering
 \subfigure[]{\label{pressure}\includegraphics[width=0.35\textwidth]{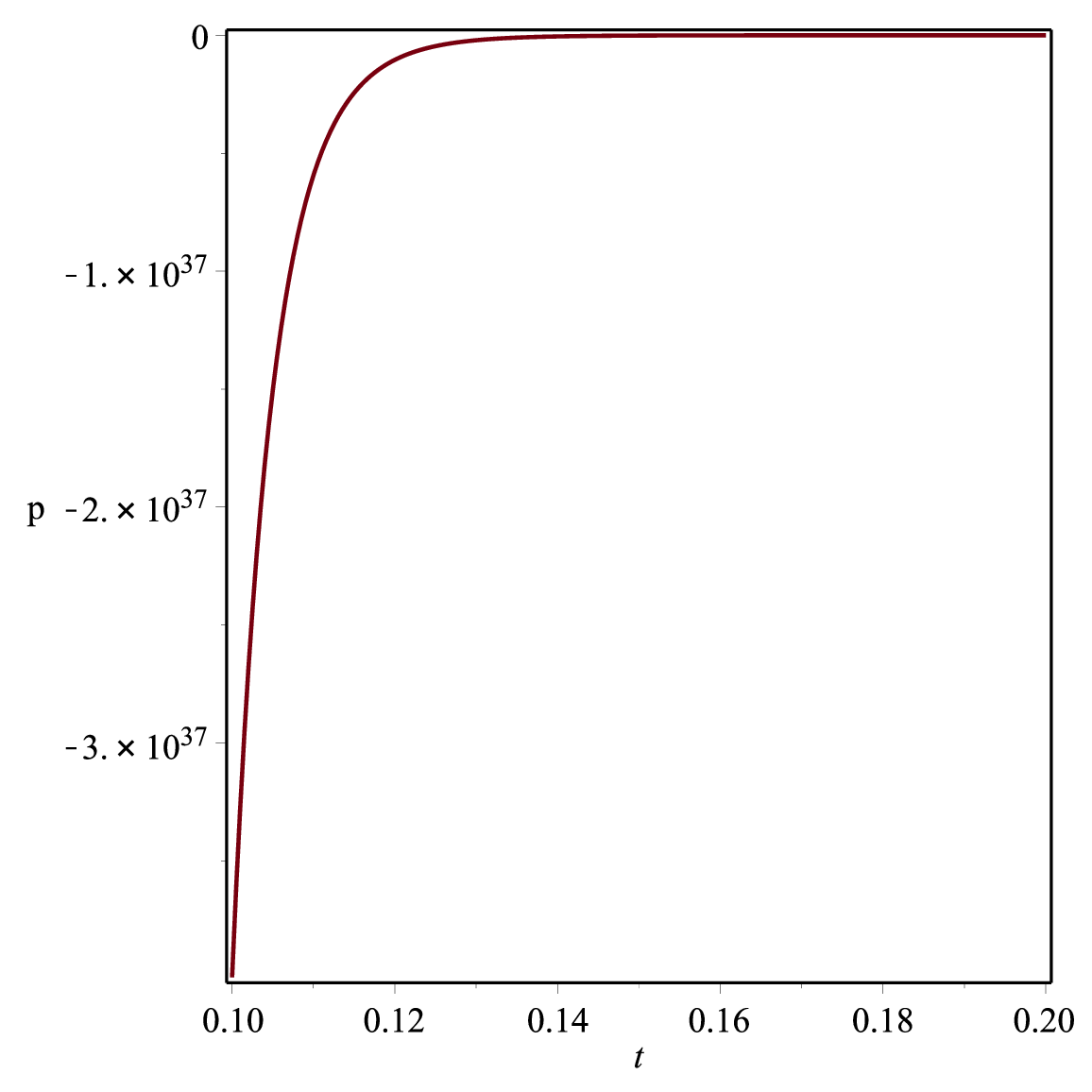}}                
 \subfigure[]{\label{rho}\includegraphics[width=0.35\textwidth]{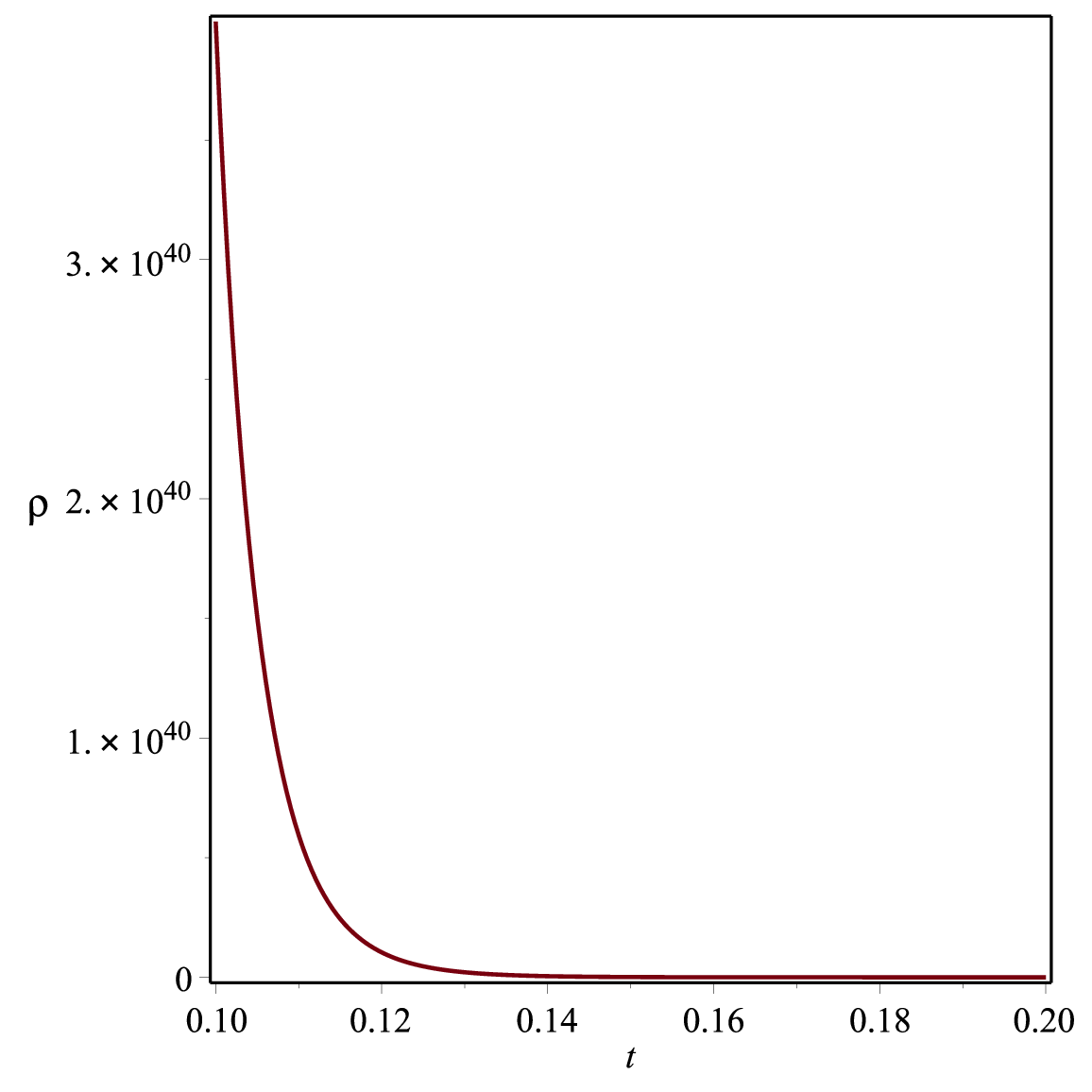}}
  \subfigure[]{\label{Lambda}\includegraphics[width=0.35\textwidth]{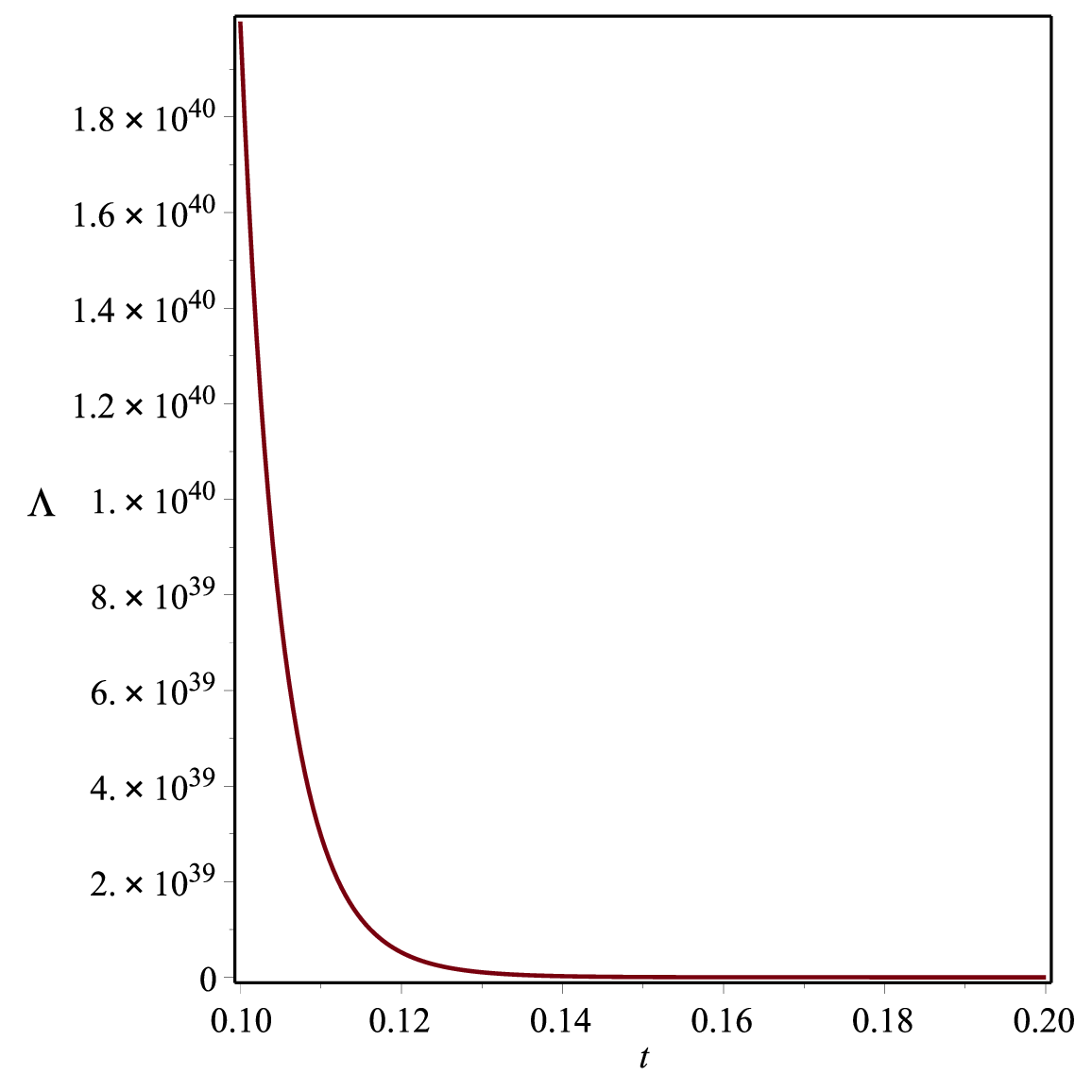}} \\
 \subfigure[]{\label{Rscalar}\includegraphics[width=0.35\textwidth]{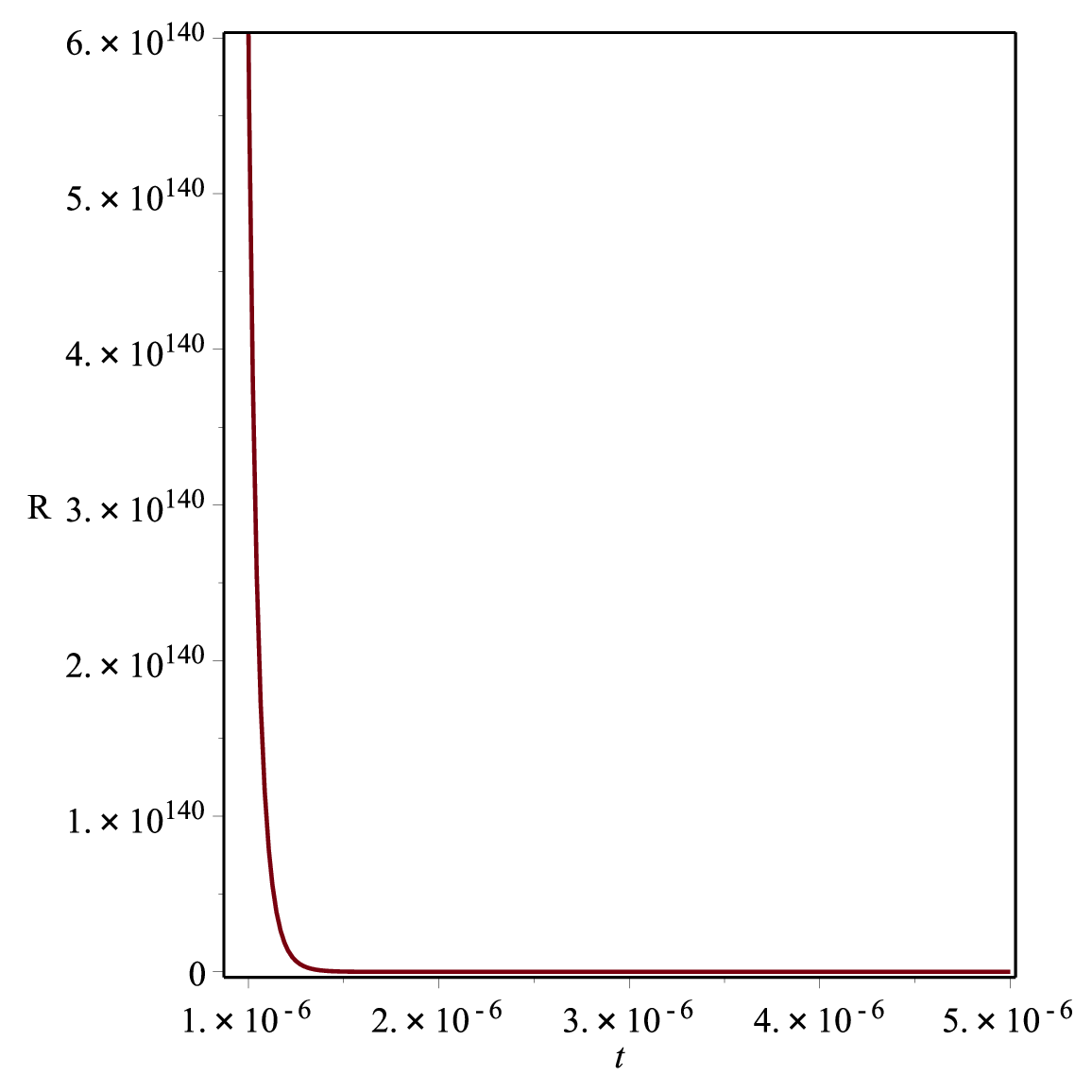}}
\subfigure[]{\label{Am}\includegraphics[width=0.35\textwidth]{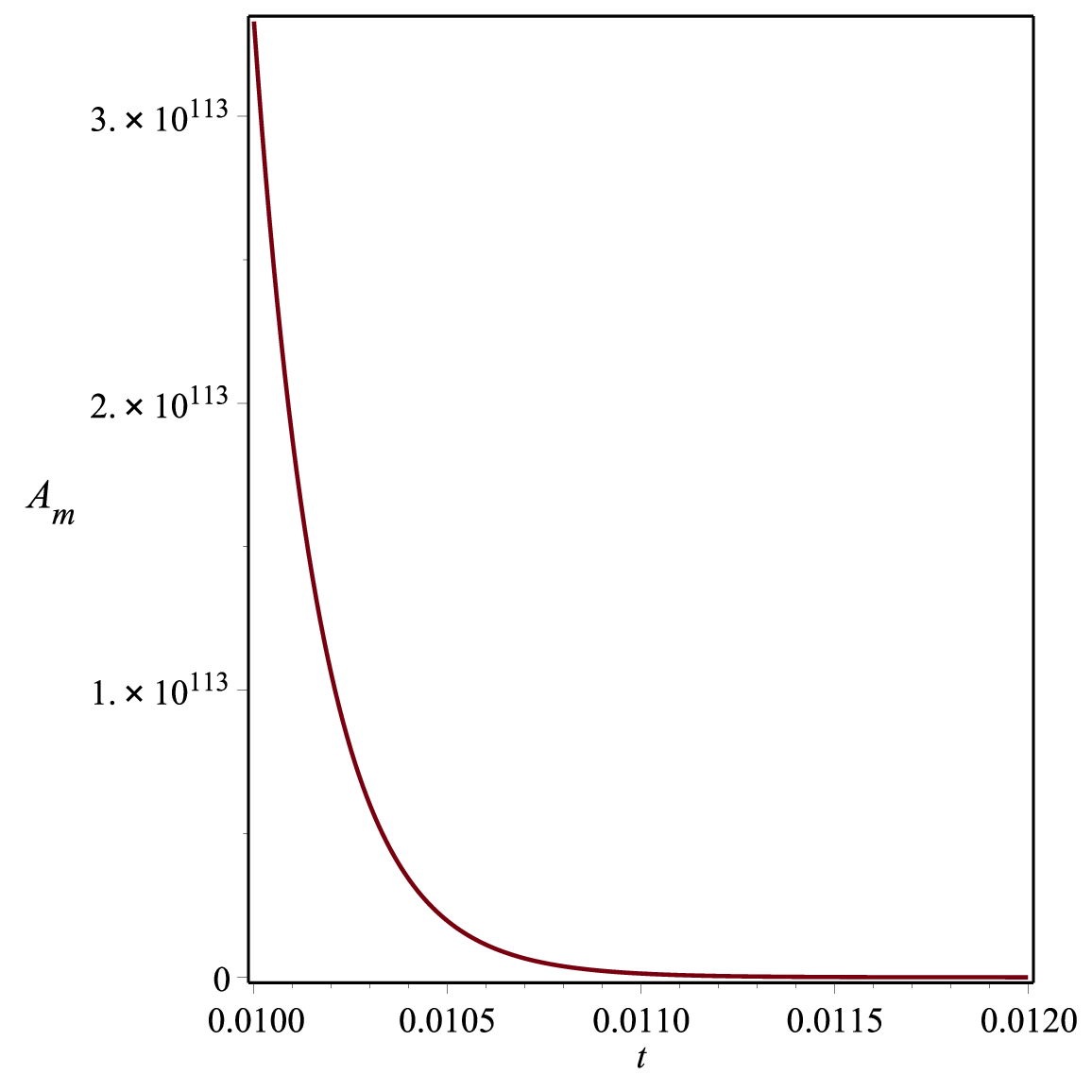}}
 \caption{Plots of $p$, $\rho$, $\Lambda$, $R$ and $A_{m}$ versus cosmic time $t$. Here $\lambda=-0.1$, $c_{1}=1.1$, $c_{2}=1$, 
$\alpha=1$ and $n=0.1$.}
 \label{prholam}
\end{figure}

\section{Physical and geometric properties of model}
Using Eqs. (\ref{eq43})$-$(\ref{eq45}) in (\ref{eq23})$-$(\ref{eq26}) and solving with (\ref{eq17}), we obtain obtain the 
expressions for pressure ($p$), energy density ($\rho$) and cosmological constant ($\Lambda$) for the model  (\ref{eq18}) as

\[
 p(t) = \frac{\lambda}{F_{1}(t)}\Bigl(C_{1}\cosh^{2}(\alpha t) + C_{2}\cosh^{4}(\alpha t) + C_{3}\cosh^{6}(\alpha t) 
+ C_{4}\cosh^{8}(\alpha t)
\]
\begin{equation}
\label{eq47}
+ C_{5}\sinh^{\frac{2(n - 1)}{n}}(\alpha t) + C_{6}\Bigr).
\end{equation}
\[
 \rho(t) = - \frac{1}{F_{1}(t)}\Bigl( K_{1}\cosh^{2}(\alpha t) + K_{2}\cosh^{4}(\alpha t) + K_{3}\cosh^{6}(\alpha t)
\]
\begin{equation}
\label{eq48}
+ K_{4}\cosh^{8}(\alpha t) + K_{5}\sinh^{\frac{2(n - 1)}{n}}(\alpha t) + K_{6}\Bigr).
\end{equation}
\begin{equation}
\label{eq49}
\Lambda =  \frac{1}{2F_{1}(t)}\left[(K_{1} - \lambda C_{1})\cosh^{2}(\alpha t) + (K_{5} - \lambda C_{5})
\sinh^{\frac{2(n - 1)}{b}}(\alpha t) + (K_{6} - C_{4})\right],
\end{equation}
where
\begin{equation}
\label{eq50}
F_{1}(t) = 16 n^{2}(8 \pi+\lambda)\sinh^{2}(\alpha t),
\end{equation}
\begin{equation}
\label{eq51}
C_{1} = 4n^{2}c_{2}^{2} + 12nc_{2}^{2},
\end{equation}
\begin{equation}
\label{eq52}
C_{2} = 18nc_{2}^{2} + 3n^{2}c_{2}^{2} + 9c_{2}^{2},
\end{equation}
\begin{equation}
\label{eq53}
C_{3} = 2n^{2}c_{2}^{2} + 18c_{2}^{2},
\end{equation}
\begin{equation}
\label{eq54}
C_{4} = - n^{2}c_{2}^{2} + 6n c_{2}^{2} + 9c_{2}^{2},
\end{equation}
\begin{equation}
\label{eq55}
C_{5} = 16n^{2},
\end{equation}
\begin{equation}
\label{eq56}
C_{6} = 4n^{2}c_{2}^{2} - 16\alpha^{2} n
\end{equation}
\begin{equation}
\label{eq57}
K_{1} = 4n^{2}c_{2}^{2}\lambda  + 12nc_{2}^{2}\lambda  - 96\lambda \alpha^{2} - 768\alpha^{2}\pi.
\end{equation}
\begin{equation}
\label{eq58}
K_{2}=18nc_{2}^{2}\lambda  + 3n^{2}c_{2}^{2}\lambda  + 9c_{2}^{2}\lambda,
\end{equation}
\begin{equation}
\label{eq59}
K_{3} = 2c_{2}^{2}\lambda n^{2} + 18c_{2}^{2}\lambda,
\end{equation}
\begin{equation}
\label{eq60}
K_{4} = -n^{2}c_{2}^{2}\lambda  + 6n c_{2}^{2}\lambda  + 9 c_{2}^{2}\lambda ,
\end{equation}
\begin{equation}
\label{eq61}
K_{5} = 16 n^{2}(32\pi + 5\lambda),
\end{equation}
\begin{equation}
\label{eq62}
K_{6} = 4n^{2}\lambda c_{2}^{2}  + 16n \lambda \alpha^{2} + 356 n \pi \alpha^{2}.
\end{equation}
Figure $1(a)$ depicts the variation of pressure versus time for $\lambda = -0.1$, $c_{1} = 1.1$, $c_{2} = 1$, 
$\alpha = 1$ and $n = 0.1$ as a representative case. From the figure we observe that pressure is increasing 
function of time. It starts from a large negative value and approaches to a small negative value near zero.
From the discovery of the accelerated expansion of the universe, it is generally assumed that this cosmic 
acceleration is due to some kind of energy-matter with negative pressure known as `dark energy'. \\

The energy density has been graphed versus time in Fig. $1(b)$. It is evident that the energy density remains 
always positive and decreasing function of time and it converges to zero as $t \to \infty$ as expected. \\

The behaviour of the universe in this model will be determined by the cosmological term $\Lambda$, this term 
has the same effect as a uniform mass density $\rho_{eff} = - \Lambda$ which is constant in time. A positive 
value of $\Lambda$ corresponds to a negative effective mass density (repulsion). Hence, we expect that in 
the universe with a positive value of $\Lambda$ the expansion will tend to accelerate whereas in the universe 
with negative value of $\Lambda$ the expansion will slow down, stop and reverse. Figure $1(c)$ is the plot of 
cosmological term $\Lambda$ versus time. From this figure, we observe that $\Lambda$ is decreasing function 
of time $t$ and it approaches a small positive value at late time (i.e. at present epoch). Recent cosmological 
observations [1$-$5, 9, 10] suggest the existence of a positive cosmological constant $\Lambda$ with the magnitude 
$\Lambda(G\hbar/c^{3})\approx 10^{-123}$. These observations on magnitude and red-shift of type Ia supernova suggest 
that our universe may be an accelerating one with induced cosmological density through the cosmological $\Lambda$-term. 
Thus, the nature of $\Lambda$ in our derived models are supported by recent observations. For detailed description 
via different theoretical models and cosmography tests, one can see the recent review by Bamba et al. \cite{ref75}.\\

The directional Hubble parameters ($H_{i}$), Hubble parameter
($H$), expansion scalar ($\theta$), spatial volume ($V$), anisotropy parameter ($A_{m}$), shear scalar
($\sigma$) and deceleration parameter ($q$) are, respectively, given by
\begin{equation}
\label{eq63}
H_{1} = \frac{\alpha}{n}\coth(\alpha t),
\end{equation}
\begin{equation}
\label{eq64}
H_{2} = \frac{\alpha}{n}\coth(\alpha t) + \frac{c_{2}}{\sinh^{\frac{3}{n}}(\alpha t)},
\end{equation}
\begin{equation}
\label{eq65} H_{3} = \frac{\alpha}{n}\coth(\alpha t) - \frac{c_{2}}{\sinh^{\frac{3}{n}}(\alpha t)},
\end{equation}
\begin{equation}
\label{eq66} \theta = 3H = \frac{3\alpha}{n}\coth(\alpha t),
\end{equation}
\begin{equation}
\label{eq67} V = ABC = (\sinh(\alpha t))^{\frac{3}{n}},
\end{equation}
\begin{equation}
\label{eq68} A_{m} = \frac{1}{3}\left(\frac{nc_{2}}{\alpha \coth(\alpha t)\sinh^{\frac{3}{n}}(\alpha t)}\right)^{2}.
\end{equation}
\begin{equation}
\label{eq69} \sigma^{2} = \left(\frac{c_{2}}{\sinh^{\frac{3}{n}}(\alpha t)}\right)^{2},
\end{equation}
\begin{equation}
\label{eq70} q = n\left(1 - \tanh^{2}(\alpha t)\right) - 1.
\end{equation}
From the  Eq. (\ref{eq70}), we observe that $q > 0$ for $t < \frac{1}{\alpha}\tanh^{-1}
(1 - \frac{1}{n})^{\frac{1}{2}}$ and $q < 0$ for $t > \frac{1}{\alpha}\tanh^{-1}(1 - \frac{1}{n})^{\frac{1}{2}}$. It is also 
observed that for for $0 < n \leq 1$, our model is in accelerating phase but for $n > 1$, our model is evolving from decelerating 
phase to accelerating phase. It follows that in our derived model, one can choose the value of $n$ which gives the physical 
behavior of DP consistent with the observations. Also, recent observations from SNe Ia, expose that the present universe 
is accelerating and the value of DP lies to some place in the range $-1 \leq q < 0$. \\

For the present Universe ($t_{0} = 13.8$ GYr) with $q_{0} = -0.73$ (Cunha \cite{ref76}), Eq. (\ref{eq70}) yields the following 
relationship between the constants $n$ and $\alpha$:
\begin{equation}
\label{eq71} \alpha = \frac{1}{13.8}\tanh^{-1}\left[1-\frac{0.27}{n}\right]
^{\frac{1}{2}}.
\end{equation}
It is self explanatory from the above relation that for the present Universe, the model is valid only for $n > 0.27$. 
It is observed that for $0 < n \leq 1$, our model is in accelerating phase but for $n > 1$, our model is evolving from decelerating 
phase to accelerating phase. It follows that in our derived model, one can choose the value of $n$ which gives the physical behavior 
of DP consistent with the observations.\\

From Eqs. (\ref{eq67}) and (\ref{eq66}), we observe that the spatial volume is zero at $t = 0$ and the expansion scalar 
is infinite, which show that the universe starts evolving with zero volume at $t = 0$ which is big bang scenario. From 
Eqs. (\ref{eq43})$-$(\ref{eq45}), we observe that the spatial scale factors are zero at the initial epoch $t = 0$ and hence 
the model has a point type singularity \cite{ref77}. We observe that proper volume increases exponentially with time. The physical 
quantities isotopic pressure ($p$), proper energy density ($\rho$), Hubble factor ($H$) and shear scalar ($\sigma$) diverge 
at $t = 0$. As $t \to \infty$, volume becomes infinite where as $p$, $\rho$, $H$, $\theta$ approach to zero. It is important 
to note here that $\lim_{t \to 0}\left(\frac{\rho}{\theta^{2}}\right)$ spread out to be constant. Therefore, the model of the 
universe goes up homogeneity and matter is dynamically negligible near the origin. This is in good agreement with the result 
already given by Collins\cite{ref78}. Finally, we can say that the model represents a shearing, non-rotating, expanding and 
accelerating universe, which starts with a big bang singularity and approaches to isotropy at the present epoch. \\

The dynamics of the mean anisotropic parameter depends on two constant $n$ and $c_{2}$ . From Eq. (\ref{eq68}), we observe that 
at late time when $t \to \infty$, $A_{m} \to 0$. Thus, our model has transition from initial anisotropy to isotropy at present 
epoch which is in good harmony with current observations. Figure $1(e)$ depicts the variation of anisotropic parameter ($A_{m}$) 
versus cosmic time $t$. From the figure, we observe that $A_{m}$ decreases with time and tends to zero as $t \to \infty$. Thus, the 
observed isotropy of the universe can be achieved in our model at present epoch. In the frame of pure $f(R)$ gravity, the pioneer 
works (Nojiri and Odintsov \cite{ref79,ref80}) present very natural unification of the early-time inflation and late-time acceleration. 
According to this observation which a is sub-case of current model there is no place for large anisotropy after inflation. But the case 
in the present model is different. In our derived model $q < 0$ for $t > t_{c}$, where the critical time $t_{c} = \frac{1}{\alpha}
\tanh^{-1}(1 - \frac{1}{n})^{\frac{1}{2}}$. Also we observe that the average scale factor $a(t) \to 0$ as $t \to 0$ and also 
$a(t) \to \infty$ as $t \to \infty$. This indicates that there exists inflation.\\

The Ricci scalar for the solution is given by
\[
 R = -\frac{1}{8n^{2}[\cosh(\alpha t)-1]}\Biggl[4\left[c_{2}^{2}(3n + n^{2}) + 24\alpha^{2}\right] \cosh^{2}(\alpha t)
\]
\[
 + 3c_{2}^{2}(n^{2} + 9 )\cosh^{4}(\alpha t) + 2c_{2}^{2}\left(n^{2} + 9 \right)\cosh^{6}(\alpha t)
\]
\begin{equation}
\label{eq72}
 - c_{2}^{2}\left(n^{2} - 6n - 9\right)\cosh^{8}(\alpha t) + 48n^{2}\sinh^{\frac{2n-2}{n}}(\alpha t) - 
 48\alpha^{2}n + 4n^{2}c_{2}^{2}\Biggr].
\end{equation}
The evolution of Ricci scalar $R$ with cosmic time is shown in Fig. $1(d)$. We observe that the curvature is positive through 
the whole evolution of the universe and $R \to 0$ as $t \to \infty$ and $R \to \infty$ when $t \to 0$ indicating the initial 
singularity. \\

\begin{figure}
 \centering
  \subfigure[$\rho-p$]{\label{energy0}\includegraphics[width=0.35\textwidth]{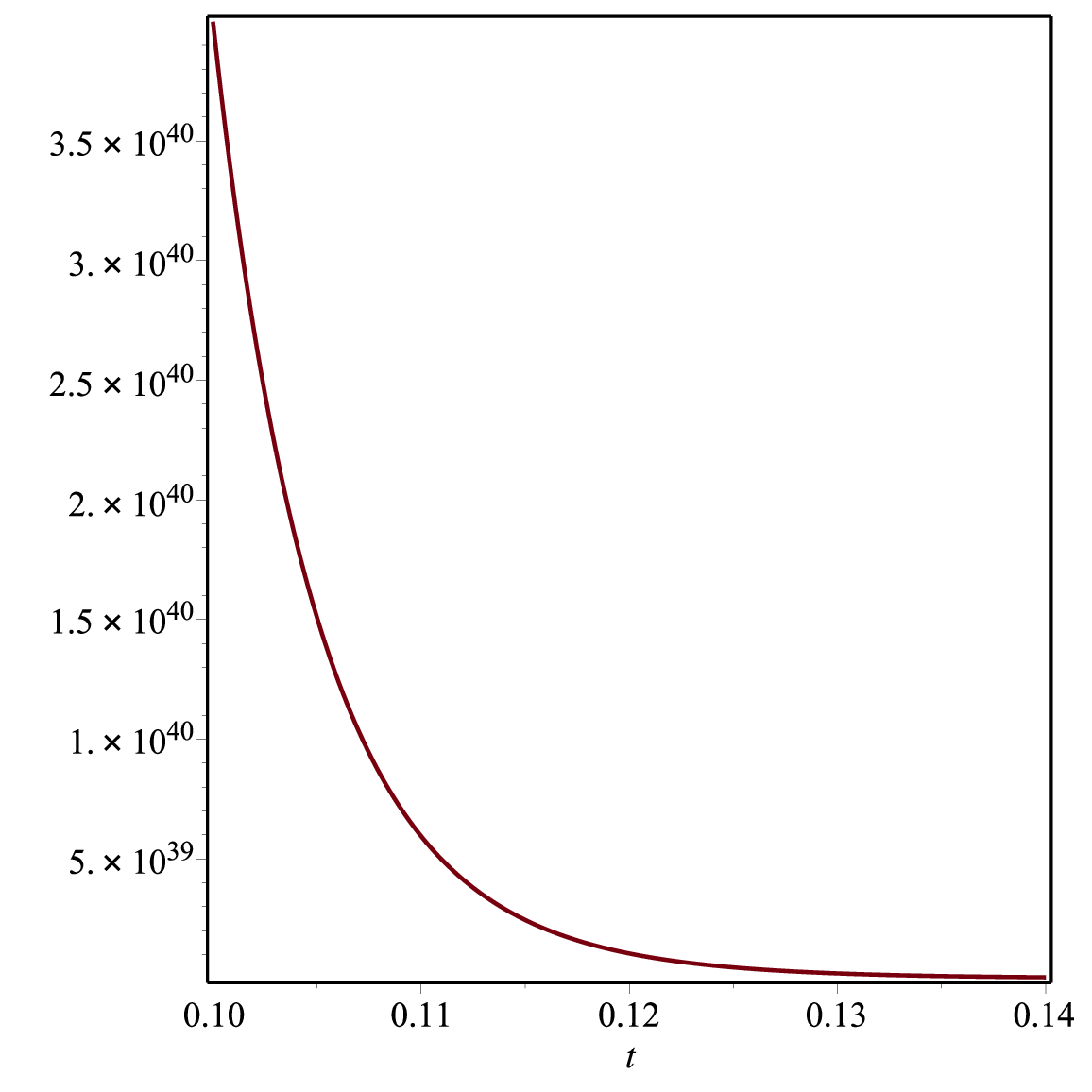}}                
  \subfigure[$\rho+p$]{\label{energy1}\includegraphics[width=0.35\textwidth]{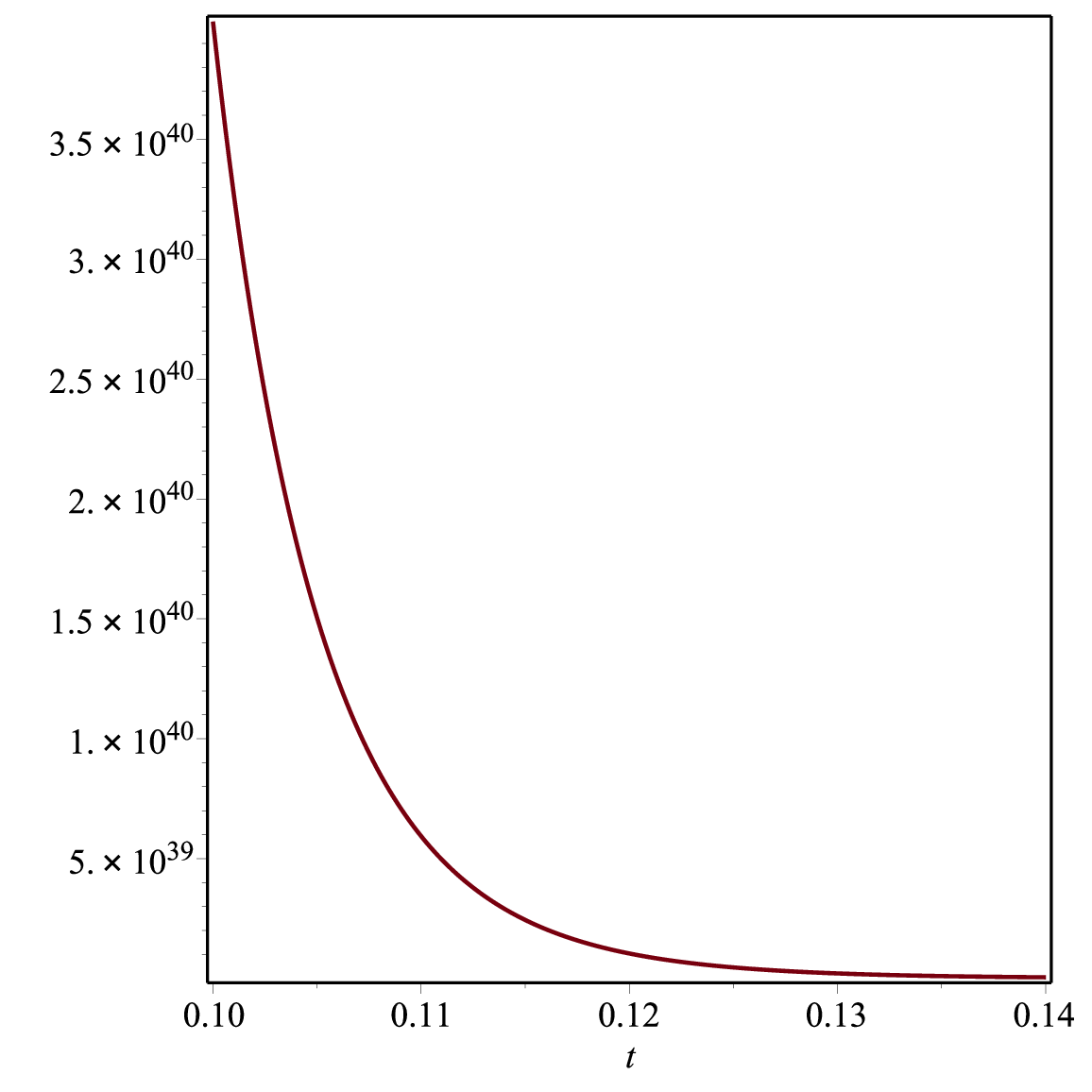}}
   \subfigure[$\rho+3p$]{\label{energy2}\includegraphics[width=0.35\textwidth]{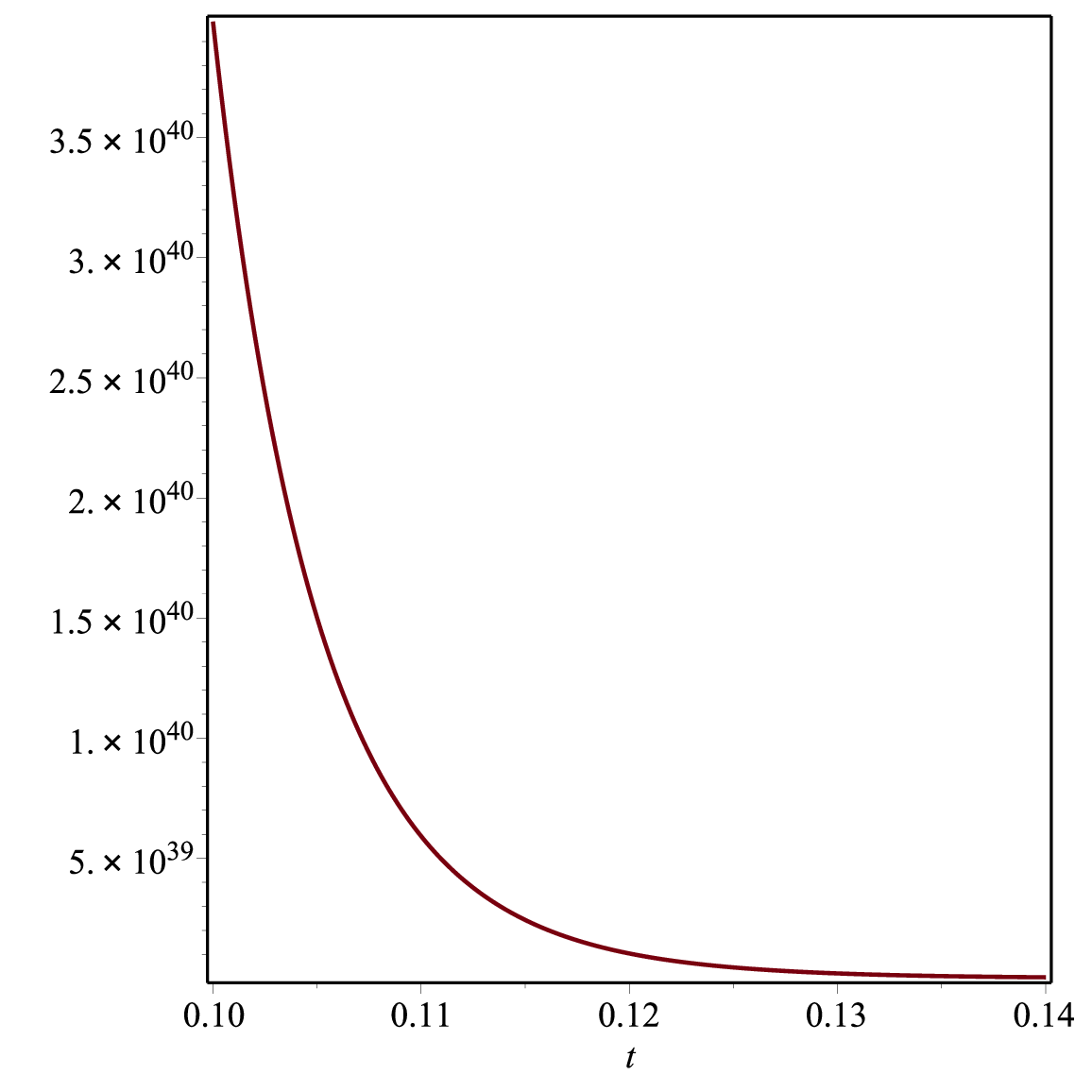}} \\
 \subfigure[$v_{s}$]{\label{sounds}\includegraphics[width=0.35\textwidth]{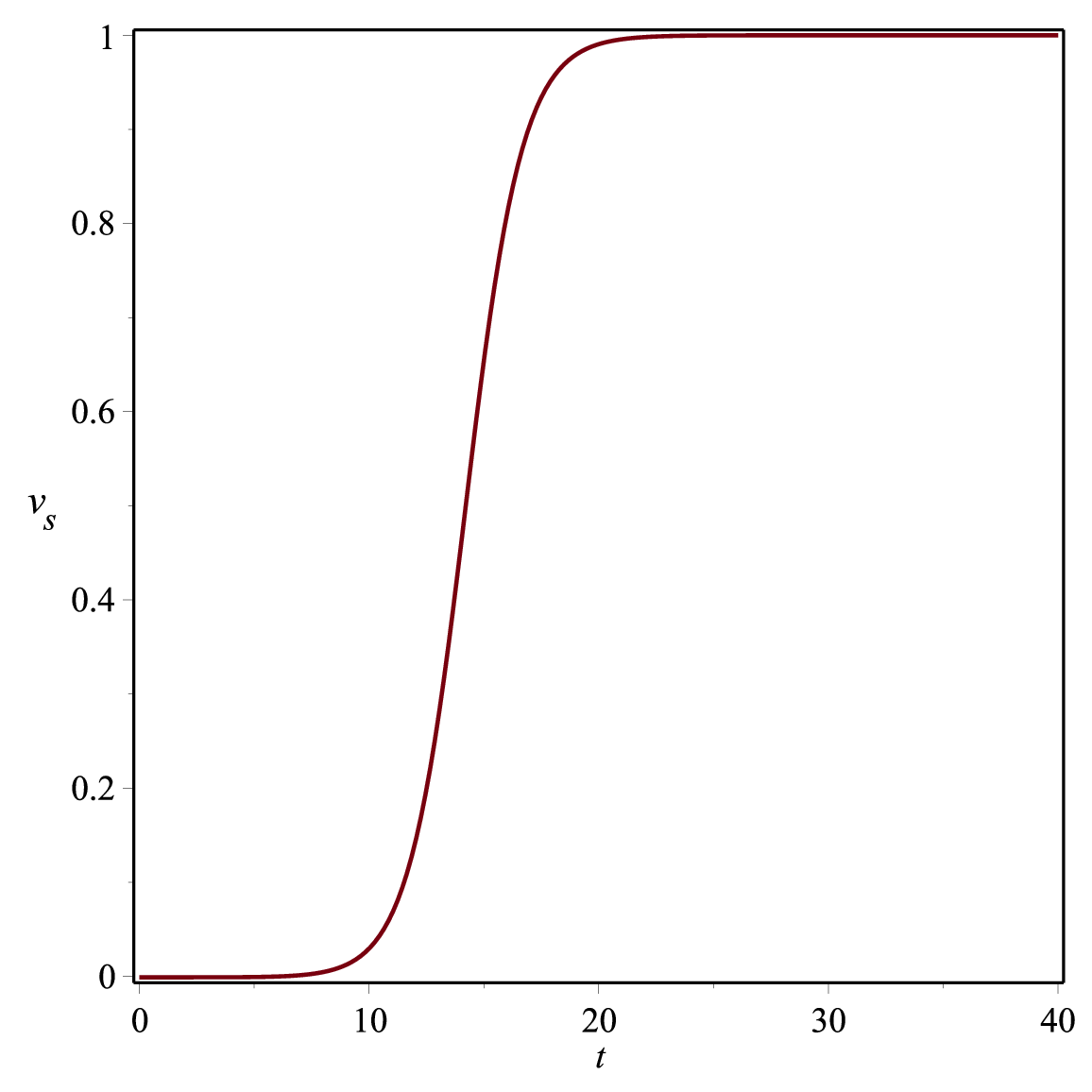}}
\subfigure[statefinder]{\label{sfinder}\includegraphics[width=0.35\textwidth]{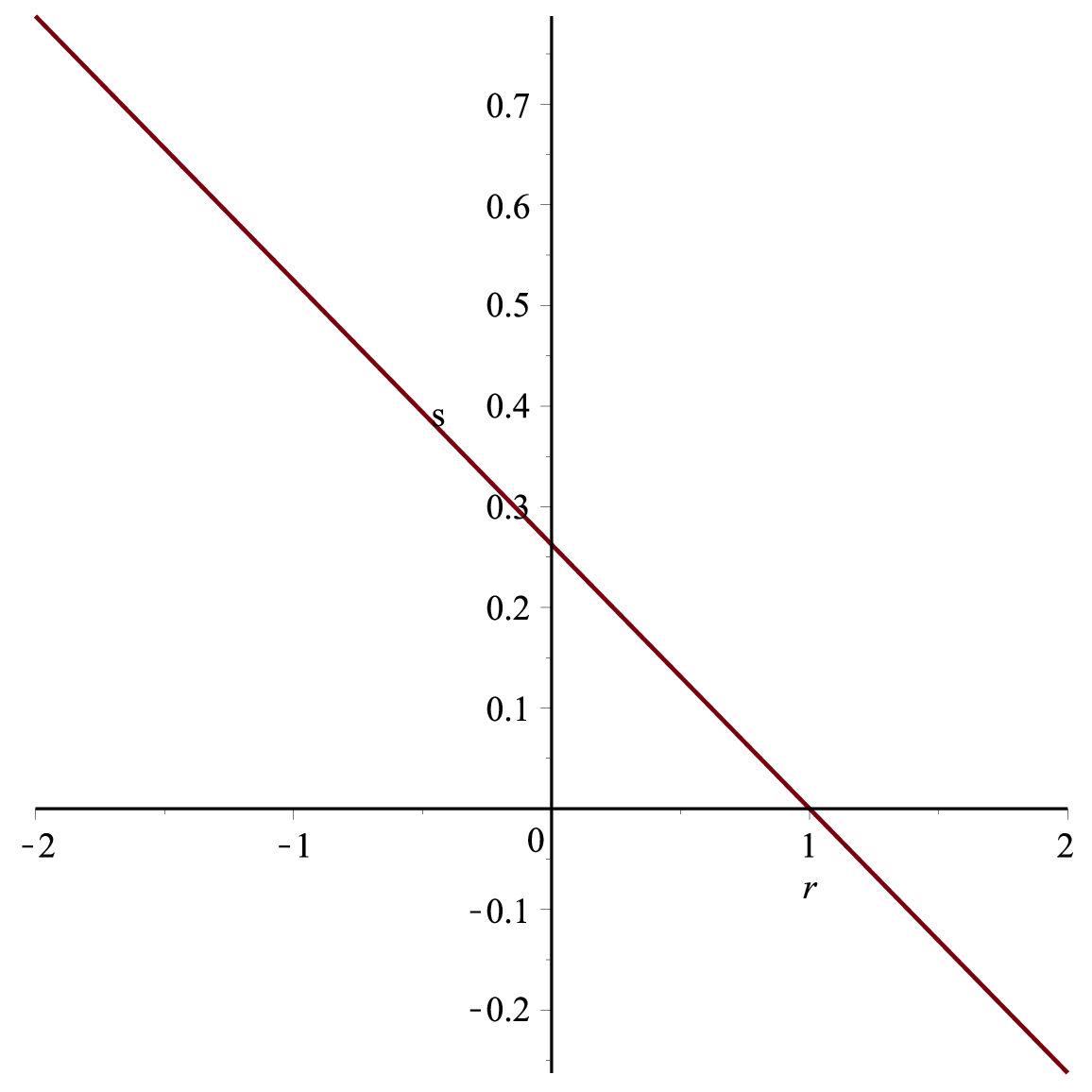}}
 \caption{The plots of energy conditions, sound speed $\upsilon_{s} $ verses time and also $s$ versus $r$. 
Here $\lambda=-0.1$, $c_{1}=1.1$, $c_{2}=1$, $\alpha=1$ and $n=0.1$.}
 \label{Energycon}
\end{figure}

\section{Physical acceptability of the solutions}
For the stability of corresponding solutions, we should check that our model is physically acceptable. \\
{\bf Sound speed:} \\
It is required that the velocity of sound $\upsilon_{s}$ should be less than velocity of light $(c)$. As we are working 
in the gravitational units with unit speed of light, i.e. the velocity of sound exists within the range $0 \leq \upsilon_{s} = 
\left(\frac{dp}{d\rho}\right) \leq 1$. \\

We obtain the sound speed as

\begin{equation}
\label{eq73}
\upsilon_{s} = \frac{dp}{d\rho}=\frac{g(t)}{h(t)},
\end{equation}
where
\[
 g(t) = \lambda \sinh^{\frac{2}{n}}(\alpha t)\Biggl[3 c_{2}^{2}(n + 3)^{2}\cosh^{8}(\alpha t) + 24 c_{2}^{2}(n + 3)\cosh^{6}(\alpha t)
\]
\[
 + 9c_{2}^{2}\left(n^{2} + 2n + 7\right) \cosh^{4}(\alpha t)  +6c_{2}^{2}\left(n^{2} + 6n + 3 \right)\cosh^{2}(\alpha t)
\]
\begin{equation}
\label{eq74}
+ 12nc_{2}^{2} - 16n\alpha^{2}\Biggr] + 16n\sinh^{2}(\alpha t).
\end{equation}
and 
\[
 h(t) = \sinh^{\frac{2}{n}}(\alpha t)\Biggl[6c_{2}^{2}\lambda (n^{2}+3)\cosh^{2}(\alpha t) + 18c_{2}^{2} \lambda \cosh^{2}(\alpha t)
\]
\[
 +9c_{2}^{2}\lambda(n^{2} + 9)\cosh^{4}(\alpha t) + (n + 3)\cosh^{6}(\alpha t) + 3c_{2}^{2}(n + 3)^{2}\lambda\cosh^{8}(\alpha t) 
\]
\begin{equation}
\label{eq75}
+ 256\alpha^{2}\pi(n-3) +16\lambda \alpha^{2}(n-6) + 512n\pi + 80n\lambda\Biggr].
\end{equation}
Here we observe that $\upsilon_{s} < 1$. Fig. $2(d)$ depicts the plot of sound speed with time. We observe that $\upsilon_{s} < 1$ 
throughout the evolution of the universe. \\
{\bf Energy conditions:} \\
The weak energy conditions (WEC) and dominant energy conditions (DEC) are given by \\

(i) $\rho \geq 0$, ~ ~ (ii) $\rho - p \geq 0$ ~ ~ and ~ ~ (iii) $\rho + p \geq 0$. \\

The strong energy conditions (SEC) are given by $\rho + 3p \geq 0$. \\

The left hand side of energy conditions have been graphed in Figs. $1(b)~\&~ 2(a, b, c)$. From these figures, we observe that 
\begin{itemize}
\item The WEC and DEC for the derived model are satisfied.  
\end{itemize}
\begin{itemize}
\item The SEC is also satisfied.
\end{itemize}
It has been shown by Wald \cite{ref81} that under very general conditions all Bianchi cosmologies (except Bianchi IX) with a 
cosmological constant and an energy momentum tensor satisfying the strong and dominant energy conditions, will unavoidably 
enter a phase of exponential expansion. With the help of this result Jensen and Stein-Schabes \cite{ref82} showed that if 
the number e-folds the Universe expands during its exponential phase is given by N then it will take a time of the order 
$ t \simeq  e^{2N}\sqrt{\Lambda}$ for anisotropy to have any effect on the observable universe. One remarkable result is 
the independence of this result from the type or magnitude of the initial anisotropy. \\

{\bf Statefinder diagnostic:} \\
Sahni et al. \cite{ref83} have introduced a pair of parameters $\{r,s\}$, called Statefinder parameters. In fact, trajectories 
in the $\{r,s\}$ plane corresponding to different cosmological models demonstrate qualitatively different behaviour. The 
statefinder parameters can effectively differentiate between different form of dark energy and provide simple diagnosis 
regarding whether a particular model fits into the basic observational data. The above statefinder diagnostic pair has 
the following form: 

\begin{equation}
\label{eq76} r = 1 + 3\frac{\dot{H}}{H^{2}} + \frac{\ddot{H}}{H^{3}} \; \; \mbox{and} \; \; 
s = \frac{r - 1}{3(q -\frac{1}{2})} \; ,
\end{equation}
to differentiate among different form of dark energy. Here $H$ is the Hubble parameter and $q$ is the deceleration parameter. 
The two parameters $\{r,s\}$ are dimensionless and are geometrical since they are derived from the cosmic scale factor $a(t)$ 
alone, though one can reproduce them in terms of the parameters of dark energy and dark matter. This pair provides information 
about dark energy in a model$-$independent way, that is, it categorizes dark energy in the context of back-ground geometry only 
which is not dependent on theory of gravity. Hence, geometrical variables are universal. \\ 

For our model, the parameters $\{r,s\}$ can be explicitly written in terms of $t$ as
\begin{equation}
\label{eq77}
r = \frac{1}{nH^{2}}\left[nH^{2} - 3\alpha^{2} csch^{2}(\alpha t) + \alpha^{3}\coth^{2}\{1 - 2csch^{2}(\alpha t)\}\right],
\end{equation}
\begin{equation}
\label{eq78}
s=\frac{-2}{3nH^{2}(2q - 1)}\left[3\alpha^{2} csch^{2}(\alpha t) - \alpha^{3}\coth^{2}\{1 + 2csch^{2}(\alpha t)\} \right].
\end{equation}
So we can get the relation between $r$ and $s$ as
\begin{equation}
\label{eq79}
s=\frac{1}{3(q-\frac{1}{2})}(r-1).
\end{equation}
Figure $2(e)$ depicts the variation of $s$ against $r$. From Fig. $2(e)$, we observe that $s$ is negative when $r \geq 1$. The figure 
shows that the universe starts from an Einstein static era ($r \to \infty, s \to - \infty$) and goes to the $\Lambda$CDM model 
($r = 1, s = 0$). \\

Therefore, on the basis of above discussions and analysis, our corresponding solutions are physically acceptable. \\ 
\section{Discussions}
Evolution of Bianchi type-$V$ cosmological model is studied in presence of perfect fluid and variable cosmological constant in 
$f(R, T)$ theory of gravity (Harko et al. \cite{ref27}). For each choice of the function $f(R,T)$, we get different theoretical 
models. Three examples have been given in Harko et al. \cite{ref27} (i) $f(R, T) = R + 2f(T)$, (ii) $f(R, T) = f_{1}(R) + f_{2}(T)$ and 
(iii) $f(R, T) = f_{1}(R) + f_{2}(R)f_{3}(T)$. The cosmological consequences of the case (i) has been discussed before by several 
authors. In this paper, the gravitational field equation has been established by taking case (ii) into consideration. To find the 
deterministic solution, we have considered a time dependent deceleration parameter which yields a scale factor as $a(t) = 
\sinh^{\frac{1}{n}}(\alpha t)$, where $n$ and $\alpha$ are positive constants. For $n > 1$, this generates a transition of the 
universe from the early decelerating phase to the recent accelerating phase. \\

The current observations of the large-scale cosmic microwave background suggest that our physical universe is expanding isotropic 
and homogeneous models with a positive cosmological constant. The analysis of CMB fluctuations may confirm this picture. But other 
analysis reveal some inconsistency. Analysis of WMAP data sets shows that the universe could have a preferred directions (Pullen et al. 
\cite{ref84}; Samal et al. \cite{ref85}; Groeneboom and Eriksen \cite{ref86}; and Armend$\acute{a}$riz-Pic$\acute{o}$n et al. \cite{ref87}). 
The ILC$-$WMAP data maps show seven axes well aligned with one another and the direction Virgo. For this reason Bianchi models are important 
in the study of anisotropies. \\ 

The main features of the models are as follows:

\begin{itemize} 
\item In summary, we considered the modified gravity which naturally unifies two expansion phases of the universe: inflation at 
early times and cosmic acceleration at current epoch.

\item The models are based on exact solutions of the $f(R,T)$ gravity field equations for the anisotropic Bianchi-V 
space-time filled with perfect fluid with time dependent $\Lambda$-term. 

\item The model represents an expanding, shearing, non-rotating and accelerating universe.

\item For suitable choice of constants the anisotropic parameter $A_{m}$ tends to zero for sufficiently
large time (Fig. $1(e)$). Hence, the present model is isotropic at late time which is consistent to the current observations.

\item It has been found that $\Lambda$ is a decreasing function of time and it converges to a small positive value at late 
time. The nature of decaying vacuum energy density $\Lambda(t)$ in our derived model is supported by recent 
cosmological observations. These observations on magnitude and red-shift of type Ia supernova suggest that our 
universe may be an accelerating one with induced cosmological density through the cosmological $\Lambda$-term.

\item In literature it is a plebeian practice to consider constant deceleration parameter. Now for a Universe 
which was decelerated in past and accelerating at present epoch, the DP must show signature flipping as already 
discussed. Therefore, our consideration of DP to be variable is physically justified. Our derived 
model is accelerating at present epoch.

\item For different choice of $n$, we can generate a class of viable cosmological models of the universe in 
Bianchi type-V space-time. For example, if we set $n = 1$ in Eq. (\ref{eq42}), we find $a = \sinh(\alpha t)$ which is 
used by Pradhan et al. \cite{ref65} in studying the accelerating dark energy models in Bianchi type-$VI_{0}$ space-time 
and Pradhan et al. \cite{ref66} in studying Bianchi type-I cosmological models with time dependent $\Lambda$-term.  It 
is observed that such models are also in good harmony with current observations.

\item We observe that our derived solutions are physically acceptable in concordance with the fulfillment of WEC, DEC and 
SEC. The sound speed $\upsilon_{s} < 1$ (see in Fig. $2(d)$). The models in $f(R,T)$ theory of gravity has stability and 
has initial singularity. Wald \cite{ref81} pointed out that Bianchi cosmologies (except Bianchi $IX$) with $\Lambda$ and 
satisfying DEC and SEC will enter in a phase of exponential expansion.   

\item $\{r,s\}$ diagram (Fig. $2(e)$) shows that the evolution of the universe starts from Einstein static era 
($r \to \infty, s \to -\infty $) and approaches to $\Lambda$CDM model ($ r = 1, s = 0$). So, from the Statefinder 
parameter $\{r,s\}$, the behaviour of different stages of the evolution of the universe have been generated.
\end{itemize}

Thus, the solutions demonstrated in this paper may be useful for better understanding of the characteristic of  
Bianchi type-V cosmological models in the evolution of the universe within the framework of $f(R, T)$ gravity theory. 

\section*{Acknowledgments}
The authors would like to thank the Inter-University Centre for Astronomy and Astrophysics (IUCAA), Pune, India for 
providing facility \& support during a visit where part of this work was done. A. Pradhan gratefully acknowledges 
the financial support by University Grants Commission, New Delhi, India under the research grant (Project F.No. 41-899/2012(SR)). 
The authors also thank Prof. V. Sahni and Prof. S. G. Ghosh for helpful discussions.

\end{document}